\documentclass[lettersize,conference,bookmarks=false]{IEEEtran}
\usepackage{amsmath,amsfonts}
\input{Format.tex}
\DeclarePairedDelimiter\parens{\lparen}{\rparen}
\DeclarePairedDelimiter\bracks{\lbrack}{\rbrack}
\DeclarePairedDelimiter\braces{\lbrace}{\rbrace}

\DeclarePairedDelimiter\norm{\lVert}{\rVert}

\DeclarePairedDelimiter\floor{\lfloor}{\rfloor}

\newcommand*{\inv}{\!^{-1}}

\DeclareMathOperator{\C}{\mathbb{C}}

\DeclareMathOperator{\Z}{\mathbb{Z}}

\makeatletter
\DeclareRobustCommand\onedot{\futurelet\@let@token\@onedot}
\def\@onedot{\ifx\@let@token.\else.\null\fi\xspace}

\def\eg{\emph{e.g}\onedot} 
\def\ie{\emph{i.e}\onedot}

\def\aka{\emph{a.k.a}\onedot}
\makeatother

\let\mathbf\bm

\newtcolorbox{stretchbox}[1][]{
    height fill,
    colback=white,
    colframe=black,
    #1
    }

\newtcolorbox{problem}[1]{
    breakable,
    toptitle=2.5pt,
    bottomtitle=2.5pt,
    colbacktitle=white,
    coltitle=black,
    colback=white,
    colframe=black,
    fonttitle=\large\bfseries,
    title={#1 \hfill Grade:\hspace*{0.15\paperwidth}\ }, %\! works as a placeholder
}

\newtcolorbox{solution}[1]{
    breakable,
    colback=black!3,
    fonttitle=\bfseries,
    title={#1},
}

\newcommand\redout{\bgroup\markoverwith{\textcolor{red}{\rule[.5ex]{2pt}{0.4pt}}}\ULon}

\allowdisplaybreaks
\title{Virtual Array for   Dual Function MIMO Radar Communication Systems using OTFS Waveforms
\thanks{The work was supported by  ARO grant W911NF2320103 and NSF grant ECCS-2320568.}
}
\author{
Kailong~Wang\,\orcidlink{0000-0002-3415-0790}, %\\
Athina~Petropulu\,\orcidlink{0000-0001-7380-7815}\\
Rutgers University, USA
}
\begin{document}
% \ninept
% \thispagestyle{empty}
% % \copyrightnotice{
% \copyright\ IEEE 2025
% % }
% % \toappear{
% To be submitted  to {IEEE ICASSP 2025}
% % }

\maketitle
\begin{abstract}
A MIMO dual-function radar communication (DFRC) system transmitting orthogonal time frequency space (OTFS) waveforms is considered. A key advantage of MIMO radar is its ability to create a virtual array, achieving higher sensing resolution than the physical receive array. In this paper, we propose a novel approach to construct a virtual array for the system under consideration.
The transmit antennas can use the Doppler-delay (DD) domain bins in a shared fashion. A number of Time-Frequency (TF) bins, referred to as private bins, are exclusively assigned to specific transmit antennas. The TF signals received on the private bins are orthogonal and thus can be used to synthesize a virtual array, which, combined with coarse knowledge of radar parameters (\ie., angle, range, and velocity), enables high-resolution estimation of those parameters.
The introduction of $N_p$ private bins necessitates a reduction in DD domain symbols, thereby reducing the data rate of each transmit antenna by 
$N_p-1$. However, even a small number of private bins is sufficient to achieve significant sensing gains with minimal communication rate loss.
\end{abstract}
%
% \begin{keywords}
% DFRC system, MIMO radar, OTFS waveforms, communication sensing tradeoff
% \end{keywords}
%
\section{Introduction}\label{sec:intro}

An emerging trend in 6G wireless systems~\cite{8869705}
% to increase spectral efficiency by providing unconstrained spectrum access to both radar and communication systems~\cite{bourdoux20206g}.
%Central to this vision 
is the integration of sensing capabilities on the wireless system hardware platform~\cite{BibEntry2024Aug}.
%
% The integration is possible in today's technology, where the RF infrastructure are identical in both radar and wireless communication systems.
% The integration is further facilitated by the frequency convergence of modern communication and sensing systems, as wireless communication systems increasingly move to higher frequency bands traditionally occupied by radar systems to access more bandwidth.
Integrated Sensing and Communication (ISAC)~\cite{ubadah2024zakotfs} devices offer efficient use of hardware and enhanced performance of both sensing and communication functions.
Dual-function radar communication systems (DFRC)~\cite{10036975} is a class of ISAC systems that use the same waveform as well as the same hardware for both sensing and communication simultaneously.
The dual use makes good utilization of bandwidth
% Combined with hardware reuse
and for this reason DFRC systems are particularly attractive in future-generation systems, such as autonomous driving vehicles and unmanned aerial vehicles (UAVs).

% \textcolor{red}{***need to discuss the advantages of MIMO in comm systems first ***MIMO systems are good for comms because...... They are also good for radar because ***** Therefore, MIMO are good candidates for DFRC****
% The spatial diversity provided by multiple antennas helps mitigate issues such as fading and interference, leading to better signal quality.}

Multiple-input multiple-output (MIMO) technology offers significant advantages for communication systems by enabling the formation of flexible beams that can track multiple targets simultaneously. When the transmit waveforms are orthogonal, MIMO radar can synthesize virtual arrays, which have a larger aperture than uniform linear arrays (ULA) with the same number of physical components, thereby improving performance~\cite{8059539}. MIMO systems with orthogonal frequency division multiplexing (OFDM) waveforms can achieve high data rates and reliable communication and have been successfully implemented in the 4G and 5G standards~\cite{8869705}. DFRC MIMO systems using OFDM waveforms have also been explored in~\cite{10036975}.

DFRC systems are envisioned to use high frequencies and will be deployed in high-mobility applications (\eg high-speed rail)~\cite{8869705}. %where OFDM systems may not work well.
In such scenarios, the introduced high Doppler shifts destroy the orthogonality of the OFDM subcarriers, resulting in Inter-Carrier Interference and performance degradation.
Moreover, in high-mobility scenarios, the channel is time-varying and cannot be accurately estimated by its time-frequency representation, making OFDM signal detection challenging.
The recently proposed Orthogonal Time Frequency Space (OTFS) waveform~\cite{7925924} overcomes the aforementioned issues.
The OTFS system employs Doppler-delay (DD) representation,
% During modulation, complex information symbols (\eg QAM) are placed on DD domain bins, and then modulated to time domain signals via a special variant of the FFT algorithm (\cref{sec:sysmodel}), which is as efficient in computation as OFDM.
under which the time-varying channels are sparsely represented and appear linear time invariant~\cite{7925924}.
This enables accurate equalization and signal detection in high-mobility scenarios.
Existing works have shown that the OTFS outperforms OFDM in high Doppler communication~\cite{7925924,hong2022delay,8424569}.
MIMO wireless systems using OTFS waveform for communication have been studied in~\cite{8647394}, where
% and for sensing in~\cite{gaudio2020hybriddigitalanalogbeamformingmimo}.
% The methods of~\cite{8647394,gaudio2020hybriddigitalanalogbeamformingmimo} give unrestricted spectral access to all transmit antennas, allowing them to share DD bins.
% \textcolor{red}{
% For symbol detection, \cite{8647394} proposes an iterative message passing algorithm, 
% It achieves a low Bit Error Rate (BER) under a high-mobility scenario, which greatly beats OFDM waveform.
% % 's $0.2$ BER at $14$dB SNR.
all antennas have unrestricted spectral access and hence the communication rate is maximized.
MIMO wireless system using OTFS waveform for sensing is studied in~\cite{gaudio2020hybriddigitalanalogbeamformingmimo}.
This work too has all antennas sharing all Delay-doppler(DD) domain bins, which leads to the transmit signals being coupled with the radar target parameters at the radar receiver.
In~\cite{gaudio2020hybriddigitalanalogbeamformingmimo}, target estimation is performed via coarse Maximum Likelihood Estimation (MLE) detection in the DD domain, followed by iterative target extraction.
Although this method efficiently removes the masking effect between targets, the MLE detector has high complexity.
%
% }
% \textcolor{green}{\cite{gaudio2020hybriddigitalanalogbeamformingmimo} discussed the complexity of MLE radar detector. Its work also simplifies radar detection but the method is different from ours.}
% Inspired by~\cite{8999605}, the proposed
MIMO ISAC with OTFS waveforms was recently considered in~\cite{keskin2023integratedsensingcommunicationsmimootfs}, where the  DD bins are assigned to transmit antennas in an exclusive manner. For a single-target scenario, this approach achieves orthogonality at the receiver,  enabling the formation of a virtual array for improved target resolution.
However,  the orthogonality would not hold in a multi-target scenario, and further, exclusive use of DD bins by antennas reduces the communication rate.

% This design achieves high sensing resolution via its ability to synthesize a virtual array. 
% \textcolor{blue}{
% \cite{keskin2023integratedsensingcommunicationsmimootfs} designs a search-track strategy to optimize sensing and communication.
% At search mode, the system in~\cite{keskin2023integratedsensingcommunicationsmimootfs} estimates target parameters until all targets can be separated.
% With only one target in each profile, the system can synthesize a virtual array with an all-exclusive design and achieves a high angle estimation resolution. \textcolor{red}{Does it take 2 transmissions to find the target? I do not get this. I need to understand this because it may also be applicable in what we do. Have you thought about this?}
% Then, the estimated parameters are used in track mode to optimize the sensing and communication beamforming.
% The sensing in search mode uses a generalized likelihood ratio test (GLRT) and achieves a larger sensing range than the conventional radar system, which is constrained by symbol duration and subcarrier spacing.
% The communication SNR is optimized by beamforming.
% }
% \textcolor{Teal}{It only proves that the all-exclusive design can synthesize a virtual array when a single target appears in the channel. It does not explain how to synthesize the VA nor prove its result holds for multiple target cases.}

In our recent work \cite{wang2025dfrcotfs},  we proposed a MIMO DFRC system that transmits OTFS waveforms and efficiently uses the available bandwidth for both radar and communication purposes.
% The system is inspired by the MIMO DFRC OFDM approach of~\cite{10036975}.
%,8059539}.
% \cite{8059539} assigns subcarriers to each transmit antenna in an exclusive fashion.
% The resulting system achieves fine radar parameters estimation but sacrifices the communication rate due to the inefficient use of the bandwidth.
% in which shared-private subcarrier allocation scheme that achieves a trade-off between the communication rate and radar parameter estimation.
% ~\cite{10036975}, there are shared subcarriers, which are available to all antennas for shared use, and also private subcarriers for exclusive use by each antenna. The shared subcarriers enable high communication rate and coarse target parameter estimation, while the private subcarriers enable finetuning of those coarse estimates via the construction of a virtual array.
% Although the shared-private subcarrier design achieves high bandwidth efficiency, the underlying OFDM waveform is not suitable for high Doppler scenarios.
In \cite{wang2025dfrcotfs},  the transmit antennas can use the DD bins in a shared fashion, which on one hand,
enables high communication rates, on the other hand, 
makes the sensing task more difficult. In \cite{wang2025dfrcotfs}, a low-complexity approach was proposed to estimate the radar parameters.
However, the target angle resolution in \cite{wang2025dfrcotfs} is limited by the size of the radar receive array.

In this paper, we consider the same system as in \cite{wang2025dfrcotfs}, and the construction of a virtual array, which along with some coarse target estimates (for example, the estimates obtained in \cite{wang2025dfrcotfs}) can significantly improve the system's sensing performance.
% \textcolor{blue}{
More specifically, we propose introducing a small number of Time-Frequency (TF) private bins, which are uniquely paired with transmit antennas.
% }%
The introduction of $N_p$ private bins necessitates a reduction in DD domain symbols, thereby reducing the data rate of each transmit antenna by $N_p-1$. 
However, even a small number of private bins is sufficient to achieve significant sensing gains with minimal communication rate loss.

% \textcolor{blue}{
% Compared to~\cite{8647394,gaudio2020hybriddigitalanalogbeamformingmimo}, we achieve the same communication and sensing performance in a DFRC system.
% Compared to~\cite{keskin2023integratedsensingcommunicationsmimootfs}, we have a minimal loss on communication rate.
% }
% \textcolor{red}{Explain how the proposed compares to [12,13,14]}
% % \textcolor{magenta}{I check the submission requirement and changed the template to conference to satisfy the requirement. Besides, do we need a new title?}

% \textcolor{red}{the channel must follow the same convention - lower case for DD, Upper case for TF}
% \vspace{-0.2in}
\section{OTFS background}\label{sec:sysmodel}
In this section we provide some OTFS background in the context of a single-input single-output communication system.
% The subscript of the antenna index is omitted.

The system transmits packet bursts, each of duration \(T=N\Delta t\) with bandwidth \(B=M\Delta f\), where \(N\) is the number of subsymbols, \(M\) is the number of subcarriers, \(\Delta t\) is the subsymbol duration, and \(\Delta f\) is the subcarrier spacing.
The orthogonality condition requires that \(\Delta t \cdot \Delta f = 1\)~\cite{mohammed2020derivation}.
In each burst, a set of \(NM\) complex symbols are arranged on the DD grid,
%\begin{equation*}
   \( \braces*{\parens*{k\Delta \nu, l\Delta\tau} \mid k=0,1,\ldots,N-1;l=0,1,\ldots,M-1},\)
%\end{equation*}
where \(k\) and \(l\) are Doppler and delay indices, respectively.
The grid spacing is \(\Delta \nu={1}/({N\Delta t})\) and \(\Delta \tau={1}/({M\Delta f})\).
The transmitted signal is reflected by $J$ point reflectors, each introducing a Doppler and delay that are integer multiples of the corresponding grid spacing, \ie \(\nu_j=k_{j}\Delta\nu\), \(k_{j}\in\Z\) and \(\tau_j=l_{j}\Delta\tau\), \(l_{j}\in\Z^{+}\).
The dimensions of the grid 
%satisfy $N\geq 2\max \abs{k_{\nu_j}}$ and $M\geq\max l_{\tau_j}$ so 
are such that can support all present Doppler and delays.
The complex gain introduced by each point is $\beta_j$.
% \textcolor{green}{We need to define $\nu_j$ and $\tau_j$}

The OTFS transmitter maps the symbol of DD domain, \ie $x[k,l]$, 
% (denoted as $X_{DD}[l+kM]$) 
to time-frequency domain, \ie $X[n,m]$, via the Inverse Symplectic Finite Fourier Transform (ISFFT)~\cite{7925924}, 
% (\ie $X_{TF}[m+nM]$ for TF bin $[n, m]$),
% \footnote{The ISFFT is $N$-point DFT along the time\textcol or{blue}{/Doppler} axis and the $M$-point IDFT along the frequency\textcolor{blue}{/delay} axis.}
\begin{equation}\label{eq:ISFFT}
\setlength{\abovedisplayskip}{1pt}
\setlength{\belowdisplayskip}{1pt}
    X[n,m] = \frac{1}{NM}\sum_{k=0}^{N-1}\sum_{l=0}^{M-1}x[k,l]e^{j2\pi\parens*{\frac{kn}{N}-\frac{ml}{M}}}.
\end{equation}
The analog signal  for transmission, \(s(t)\), is created via the  Heisenberg Transform~\cite{7925924}, \ie
\begin{equation}\label{eq:Heisenberg} 
\setlength{\abovedisplayskip}{1pt}
\setlength{\belowdisplayskip}{1pt}
    % \( 
    s(t) = \sum_{n=0}^{N-1}\sum_{m=0}^{M-1}X[n,m]g_{tx}(t-n\Delta t)e^{j2\pi m\Delta f(t-n\Delta t)}
    % \)
    ,
\end{equation}
where \(g_{tx}(t)\) is the pulse function of the transmitter.
% In the rest of the paper, we assume all pulse functions are a rectangular pulse.
The baseband signal, \(s^{k,l}(t)\), corresponding to $x[k,l]$ is~\cite{ubadah2024zakotfs} 
% \textcolor{blue}{\cite[eq 103]{ubadah2024zakotfs}. It is $\delta$ for the case of infinity bandwidth and $\sinc$ for realistic bandlimited signal. Can also be seen from~\cite[eq 14 15]{mohammed2020derivation}}
% \textcolor{purple}{\Cref{eq:ISFFT} is $\mathbf{X}=(\mathbf{F}_N^H \otimes \mathbf{F}_M) \mathbf{x}$. \Cref{eq:Heisenberg} is $s(t)=(\mathbf{G}_{tx}\otimes \mathbf{F}_M^H)\mathbf{X}$. Then~\cref{eq:baseband} is $s(t)=(\mathbf{G}_{tx}\otimes \mathbf{F}_M^H)(\mathbf{F}_N^H \otimes \mathbf{F}_M) \mathbf{x}=(\mathbf{G}_{tx}\otimes \mathbf{F}_N^H)\mathbf{x}$. Then $g_{tx}$ is $\delta$ for ideal case or $\sinc$ for bandlimited case in~\cref{eq:baseband}. The $e^{j2\pi\frac{nk}{N}}$ is the realization of $\mathbf{F}_N^H$. Maybe I should just use $g_{tx}$.}
\begin{equation}\label{eq:baseband}
\setlength{\abovedisplayskip}{1pt}
\setlength{\belowdisplayskip}{1pt}
    s^{k,l}(t)=\sum_{n=0}^{N-1}x[k,l]e^{j2\pi\frac{nk}{N}}g_{tx}(t-n\Delta t-\frac{l}{M}\Delta t). 
\end{equation}
% and can be viewed as the Zak Transform of $d[l+kM]$, which is a composition of an ISFFT and a Heisenberg Transform.
Omitting noise, the analog received signal is 
\textcolor{blue}{
}
% (\cref{eq:channel}).
% The noiseless received signal 
{
\setlength{\abovedisplayskip}{1pt}
\setlength{\belowdisplayskip}{1pt}
\begin{equation}\label{eq:received}
    r(t)=\sum_{j=0}^{J-1} \beta_j s(t-\tau_j)e^{j2\pi\nu_j (t-\tau_j)}.
\end{equation}
}%

At the receiver, upon matched filtering and sampling
% a matched filter computes the cross-ambiguity \(A_{g_{rx},y}(t,f)\) between received signals \(y(t)\) and the receiver pulse function \(g_{rx}(t)\),
% \begin{equation}
%     \label{eq:CAF}
%     Y(t,f)=A_{g_{rx},y}(t,f)\stackrel{\Delta}{=}\int y(t')g_{rx}\!^*(t'-t)e^{-j2\pi f(t'-t)}\diff t'.\notag
% \end{equation}
% By sampling \(Y(t,f)\)
for a duration $T$ at frequency $B$ with receiver pulse function $g_{rx}(t)$ (\aka Wigner Transform~\cite{7925924}), the time-frequency domain samples \(Y[n,m]\) are obtained.
Assuming that \(g_{tx}\), \(g_{rx}\) are bi-orthogonal,
it holds that
% \begin{equation}\label{eq:Wigner}
%     Y_{TF}[m+nM]
%     % =Y(t,f)\Big|_{t=n\Delta t,f=m\Delta f} %\notag\\
%     %\stackrel{\Delta}{=}
%     =\sum_{n'=0}^{N-1}\sum_{m'=0}^{M-1}X_{TF}[n',m']H_{n,m}[n',m'],
% \end{equation}
% \textcolor{red}{define that channel}
% sample of circular convolution of~\cref{eq:channel} with \(A_{g_{rx}g_{tx}}[n,m]\).
% When \(g_{tx}\), \(g_{rx}\) are bi-orthogonal 
%\ie
% (\(
%     A_{g_{rx},g_{tx}}(t,f)\Big|_{t=n\Delta t,f=m\Delta f}
%     % =\int g_{rx}(t')g_{tx}\!^*(t'-t)e^{-j2\pi f(t'-t)}\diff t'
%     =\delta[n]\delta[m]
% \)), and
% \cref{eq:Wigner} becomes
% {
% \setlength{\abovedisplayskip}{1pt}
% \setlength{\belowdisplayskip}{1pt}
\begin{align}
    Y[n,m]
    &=X[n,m]H[n,m], \label{eq:TFIO} \\
    H[n,m]
    &=\sum_{j=0}^{J-1} \beta_j e^{-j2\pi\nu_j\tau_j} e^{j2\pi(\nu_j n\Delta t-m\Delta f\tau_j)}. \label{eq:TFChannel}
\end{align}
% }%
% where \(H\) is the effective time-frequency channel~\cite{7925924,8424569,wei2020transmitterreceiverwindowdesigns}.
% \textcolor{red}{This sections about communication system. Explain how the transmitted symbols can be receiver at the receiver. *********}
% where \(H_{n,m}\) is the time-frequency channel.
% \(H\) can be viewed as the circular convolution of the ISFFT of~\cref{eq:channel} with \(A_{g_{rx},g_{tx}}(t,f)\Big|_{t=n\Delta t,f=m\Delta f}\)~\cite{8424569}.
%
% The~\cref{eq:Wigner} is Wigner Transform~\cite{7925924}.
%By applying an SFFT on \(Y_{TF}\), we obtain the DD samples,
% \begin{equation}
%     \label{eq:SFFT}
%     Y_{DD}[k,l]=\sum_{n=0}^{N-1}\sum_{m=0}^{M-1}Y_{TF}[n,m]e^{-j2\pi\parens*{\frac{kn}{N}-\frac{ml}{M}}}.
% \end{equation}
% \textcolor{red}{
% Assuming that the channel can be estimated via the use of pilots, based on \cref{recovery}, \(X_{TF}[n,m]\) can be obtained and lead to \(X_{DD}[n,m]\) via an SFFT, i.e., 
% \begin{equation}
%     \label{eq:SFFT}
%     X_{DD}[k,l]=\sum_{n=0}^{N-1}\sum_{m=0}^{M-1}X_{TF}[n,m]e^{-j2\pi\parens*{\frac{kn}{N}-\frac{ml}{M}}}.
% \end{equation}
% *** is this ok?}
% \textcolor{green}{It's mathematically correct. But the receiver won't be able to estimate the $H$ which is in TF domain. The OTFS works in DD domain.}
% \textcolor{red}{The symbols \(X_{DD}[k,l]\) can be recovered as follows *****explain}
% \subsection{OTFS Input-Output Relation (I/O)}\label{sec:OTFS_IO}
% \textcolor{red}{
 The received DD domain symbols, \(y[k,l]\), can be obtained by applying an  SFFT on \(Y[n,m]\). It holds that as~\cite{7925924}
% {\small
\begin{align}
    y[k,l]
    % &= \sum_{k'=0}^{N-1}\sum_{l'=0}^{M-1}X_{DD}[k',l']h[k-k',l-l'] \notag\\
    % &= \sum_{j=0}^{J-1}\beta_j e^{-j2\pi\frac{k_{\nu_j}l_{\tau_j}}{NM}} X_{DD}[[k-k_{\nu_j}]_N, [l-l_{\tau_j}]_M] \label{eq:OTFS_2}\\
    % &= \sum_{k'=0}^{N-1}\sum_{l'=0}^{M-1}e^{j2\pi\frac{(k-k_{\nu_j})l_{\tau_j}}{NM}}X_{DD}[k-k',l-l']h[k',l']
    &= 
    % \sum_{j=0}^{J-1}
    \sum_{k'=0}^{N-1}\sum_{l'=0}^{M-1}x[k-k', l-l']h[k',l'], \label{eq:OTFS} \\
% \end{equation}
% }
% \textcolor{red}{
% where \(h[\cdot,\cdot]\) is the DD domain  channel
% }
% given as the SFFT of $H[\cdot,\cdot]$.
% Under the integer Doppler, bi-orthogonal pulses and rectangular windowing function assumptions, \(h[\cdot,\cdot]\) can be simplified as~\cite{wei2020transmitterreceiverwindowdesigns}
% \begin{equation}
% \setlength{\abovedisplayskip}{1pt}
% \setlength{\belowdisplayskip}{1pt}
    h[k,l]
    &=
    \sum_{j=0}^{J-1}
    \beta_j e^{-j2\pi\frac{k_{j}l_{j}}{NM}} \delta[k-k_{j}]\delta[l-l_{j}]. \label{eq:channel}
\end{align}
Substituting eq.~\eqref{eq:channel} into eq.~\eqref{eq:OTFS} yields~\cite{8424569}
% {\small
\begin{equation}\label{eq:IO}
\setlength{\abovedisplayskip}{1pt}
\setlength{\belowdisplayskip}{1pt}
    y[k,l]
    = \sum_{j=0}^{J-1}\beta_je^{-j2\pi\frac{k_{j}l_{j}}{NM}}x[[k-k_{j}]_N,[l-l_{j}]_M],
\end{equation}
% }%
where $[\cdot]_N$ is modular operation with order $N$.

Eq~\eqref{eq:IO} can be vectorized as~\cite{8424569}
% {\small
\begin{equation}\label{eq:vector_io} 
\setlength{\abovedisplayskip}{1pt}
\setlength{\belowdisplayskip}{1pt}
    \mathbf{y}
    =  \mathbf{h}\mathbf{x}+\mathbf{w}, 
\end{equation}
% }%
where $\mathbf{y}$, $\mathbf{x}$, are DD domain vectorized received symbols and transmitted symbols whose ($l+kM$)-th element is $y[k,l]$ and $x[k,l]$, respectively; 
$\mathbf{w}\in\C^{NM}$ is noise,
% additive white Gaussian noise (AWGN) with power $\sigma_{\mathbf{w}}^2$.
% $\mathbf{F}_{\textnormal{N}}$ is $N$-point DFT matrix.
% $\mathbf{I}_{\textnormal{N}}$ is $N\times N$ identity matrix representing rectangular pulses and windowing functions.
% $\mathbf{H}_t$ is the time domain channel taking effect in~\cref{eq:received}.
% % $\mathbf{H}_{TF}$ is the matrix representation of~\cref{eq:TFChannel}.
% \textcolor{blue}{
$\mathbf{h}\in\C^{NM\times NM}$ is the  channel matrix.
% }
% $\mathbf{H}$ can be estimated with a single pilot~\cite{hong2022delay}, making OTFS more efficient in channel estimation than OFDM ,which requires a full grid of pilots in time-varying channels.
$\mathbf{h}$ can be estimated with a single pilot~\cite{hong2022delay}, and the information bearing symbols   can be estimated  via LMMSE equalization.
% is
% \begin{align}\label{eq:lmmse}
%     \tilde{{\mathbf{d}}}
%     =(\mathbf{H}^{\mathsf{H}}\mathbf{H}+\sigma_{\mathbf{w}}^2\mathbf{I})^{-1}\mathbf{H}^{\mathsf{H}}\hat{\mathbf{d}}.
% \end{align}
% Then the symbol can be detected by conventional detectors, \eg correlation detectors, \etc.

% \bigskip
% \textcolor{magenta}{**** the following refers to background on OTFS for sensing. Who has proposed the method that you describe in the next 2 lines????  *****}
% \textcolor{red}{Since $H[\cdot,\cdot]$ contains Doppler and delay information of targets, the indices of \(h\)'s peak can be used to estimate the target's range and velocity.~\cite{9764285,zhang2023radarsensingotfssignaling}}
% \textcolor{blue}{This is addressed in~\cref{eq:channel}}

\section{MIMO DFRC with all DD bins used as shared }\label{sec:target}
{% \tiny
\begin{table}[H]
    \centering
    \caption{List of Variable Notations}
    \label{tab:notation}
    \resizebox{\columnwidth}{!}{%
    \begin{tabular}{|l||l|}
    \hline
    $x_{n_t}[k,l]$    & DD transmit symbol on bin $[k,l]$ at antenna $n_t$  \\ \hline
    $X_{n_t}[n,m]$    & TF transmit symbol on bin $[n,m]$ at antenna $n_t$   \\ \hline
    $s_{n_t}(t)$      & Baseband transmit signal at antenna $n_t$         \\ \hline
    $s_{n_t}^{k,l}(t)$& Baseband transmit signal of symbol on DD bin $[k,l]$   \\ \hline
    $r_{n_r}(t)$      & Baseband receive signal at antenna $n_r$          \\ \hline
    $r_{n_r}^{k,l}(t)$& Baseband receive signal of symbol on DD bin $[k,l]$   \\ \hline
    $y_{n_r}[k,l]$    & DD receive symbol on bin $[k,l]$ at antenna $n_r$   \\ \hline
    $Y_{n_r}[n,m]$    & TF receive symbol on bin $[n,m]$ at antenna $n_r$   \\ \hline
    $h[k,l]$& DD domain effective channel of bin $[k,l]$   \\ \hline
    $H[n,m]$& TF domain effective channel of bin $[n,m]$  \\ \hline
    $\mathbf{x}_{n_t}$/$\mathbf{y}_{n_r}$& Vectorized transmit/receive DD domain symbols.   \\ \hline
    $\mathbf{X}_{n_t}$/$\mathbf{Y}_{n_r}$& Vectorized transmit/receive TF domain symbols.   \\ \hline
    $\mathbf{h}$/$\mathbf{H}$& Vectorized DD/TF channel representation.   \\ \hline
    $h^j[k,l]$/$H^j[n,m]$& Effective channel of target $j$ \\ \hline
    $\omega_j$  & Spatial frequency of target $j$   \\  \hline
    $A_j[k,l]$  & DD angle profile of target $j$ on bin $[k,l]$ \\ \hline
    $A_j'[k,l]$  & Estimated DD angle profile of target $j$ on bin $[k,l]$ \\ \hline
    \end{tabular}%
    }
\end{table}
}%

Here, we outline the approach of \cite{wang2025dfrcotfs} for obtaining limited resolution target estimates.
We consider a DFRC system comprising a monostatic MIMO radar with \(N_t\) transmit antennas, \(N_r\) receive antennas and an \(N_c\) antenna communication receiver.
The carrier frequency is \(f_c\) Hz, \ie, the wavelength is $\lambda=c/f_c$ with $c$ being the speed of light.
The transmit and receive antennas form uniform linear arrays (ULA) with spacing \(g_t\) and \(g_r\), respectively.
Due to the multiple transmit and receive antennas, all signals will be indexed by the operating antenna.
The notation used in this section is summarized in table~\ref{tab:notation}.

% The radar transmits an OTFS waveform carrying information to an 
% % \textcolor{red}{
% \(N_c\)-antenna 
% % }
% communication receiver.
At the transmitter, the modulated binary source data are divided into \(N_t\) parallel streams, one for each transmit antenna.
Each antenna constructs and transmits an OTFS waveform.
We assume that 
all \(NM\) DD and TF bins are available to all transmit antennas to transmit on.
Let \(x_{n_t}[k,l]\) be the symbol that the $n_t$-th antenna places on the DD bin \([k,l]\). 
Suppose there are $J_T$ targets in the transmitter far-field, and let $\theta_j\in\bracks*{-\frac{\pi}{2},\frac{\pi}{2}}$, \(\nu_j,\tau_j\) be the  angle, Doppler and delay corresponding to target $j$. 
% \textcolor{red}{
We assume that the targets fall on the DD grid.
% so \(k_j,l_j\) are integers.
% }
%
The baseband equivalent of the signal received by the $n_r$-th receive antenna corresponding to bin $[k,l]$ is
%for $n_r=0,1,\ldots,N_r-1$, is
{
\begin{multline}\label{eq:ReceivedSignal} 
\setlength{\abovedisplayskip}{1pt}
\setlength{\belowdisplayskip}{1pt}
    r_{n_r}^{k,l}(t)
    = \sum_{j=0}^{J_T-1} \sum_{n_t=0}^{N_t-1} e^{-j2\pi (n_r g_r+n_t g_t)\frac{\sin(\theta_j)}{\lambda}}
    % \notag\\
    % &\times 
    % \sqrt{\Delta \tau}
    \sum_{n=0}^{N-1} x_{n_t}[k,l] e^{j2\pi \frac{nk}{N}} \\
    \times \beta_j g_{tx}(t-n\Delta t-\frac{l}{M}\Delta t-\tau_j)e^{j2\pi \nu_{j}(t-\tau_j)}.
\end{multline}
}%
In the above, the radar parameters, \(\theta_j, \nu_j, \tau_j\), are coupled with the transmitted symbols.
While one could use a MLE approach to obtain the radar parameters, this would involve high complexity. 
Next, we propose a low-complexity estimation approach suitable for practical implementation. 

\medskip
\noindent\underline{Target angle estimation} - 
% Let us express the continuous time $t$ as $t=\mu T+\tau$ where $\tau\in[0, T)$. 
The demodulated symbol of the $n_r$-th receive antenna corresponding to bin $[k,l]$ is (see eq~\eqref{eq:OTFS})
{ 
\begin{align}
    y_{n_r}[k,l]
    &= 
    \sum_{j=0}^{J_T-1} 
    \sum_{n_t=0}^{N_t-1} e^{-j2\pi (n_rg_r+n_tg_t)\frac{\sin(\theta_j)}{\lambda}}\sum_{k'=0}^{N-1}\sum_{l'=0}^{M-1} \notag \\
    \times &x_{n_t}[k-k',l-l']h^j[k',l'], \label{eq:DemodulatedSymbol} 
    \\
    % = \sum_{j=0}^{J_T-1}\sum_{n_t=0}^{N_t-1} e^{-j2\pi (n_rg_r+n_tg_t)\frac{\sin(\theta_j)}{\lambda}} \\
    % \times e^{-j2\pi\frac{k_{\nu_j}l_{\tau_j}}{NM}}d[n_t, [l-l_{\tau_j}]_M+[k-k_{\nu_j}]_NM].
    h^j[k,l]
    &= \beta_j e^{-j2\pi\frac{k_{{j}}l_{{j}}}{NM}} \delta[k-k_{{j}}]\delta[l-l_{{j}}]. \label{eq:anglechannelprofile}
\end{align}
}%
In the above, \(h^j [k,l]\) is DD domain sensing channel corresponding to the \(j\)-th target.
On lumping into \(A_j[k,l]\) all terms of eq.~\eqref{eq:DemodulatedSymbol} that do not depend on $n_r$,   
% \textcolor{blue}{}
we can rewrite eq.~\eqref{eq:DemodulatedSymbol} as
\begin{equation}
\setlength{\abovedisplayskip}{1pt}
\setlength{\belowdisplayskip}{1pt}
    y_{n_r}[k,l]
    =\sum_{j=0}^{J_T-1}A_j[k,l]e^{-j2\pi n_r\omega_j},  \ \ \omega_j=g_r{\sin(\theta_j)}/{\lambda}\label{eq:DemodulatedSymbol2} 
    % A_j[k,l]
    % &=\sum_{n_t=0}^{N_t-1} e^{-j2\pi n_tg_t\frac{\sin(\theta_j)}{\lambda}} \notag\\
    % \times \beta_j e^{-j2\pi\frac{(k-k_{\nu_j})l_{\tau_j}}{NM}} &d[n_t, l+kM]h[k_{\nu_j}-k, l_{\tau_j}-l],
    % \label{eq:Amplitude}\\
\end{equation}
where 
% \(\omega_j=g_r{\sin(\theta_j)}/{\lambda}\) and
{\small
\begin{equation}
    A_j[k,l]
    = \sum_{n_t=0}^{N_t-1} e^{-j2\pi n_tg_t\frac{\sin(\theta_j)}{\lambda}}\sum_{k'=0}^{N-1}\sum_{l'=0}^{M-1}
    % \notag\\
    % &\times 
    x_{n_t}[k-k',l-l']h^j[k',l']. \label{eq:AmplitudeDecomposition} 
\end{equation}
}%
% \textcolor{red}{what is $h$??}

% The symbol $d[\cdot,\cdot]$ is known to the radar receiver as this is a monostatic radar system.
On assuming that $N_r>J_T$, 
% For fixed $k$, $l$, based on\cref{eq:DemodulatedSymbol2},
\(\{y_{n_r}[k,l]\mid n_r=0,1,\ldots,N_r-1\}\) can be viewed as
the sum of $J_T$ complex sinusoids with spatial frequencies $\omega_j(k,l)$ and  complex amplitudes $A_j[k,l]$.
We can estimate the sinusoid frequencies, and thus $\theta_j$, via an   an  $N_r$ point DFT  applied on that sequence.
% and estimate the target angles as
% \begin{equation}\label{eq:AngleEstimation}
%     \theta_j = \arcsin\parens*{\lambda{\omega(j,l+kM)}/{g_r}}.
% \end{equation}
% \textcolor{red}{
% While $A_j[k,l]$ can be small for some $k,l$, in which case the peaks in the DFT will be missed, the estimation can be repeated on all $NM$ bins.
The estimation can be repeated across all $NM$ bins, and this diversity helps achieve high-quality target angle estimates.

\medskip
\noindent\underline{Target range and velocity estimation - } 
% \textcolor{blue}{
Since the resolution of the DFT is $\pi/N_r$, if $N_r$ is small there may be multiple targets corresponding to the estimated angle $\theta_j$.
If there are $N_j$ targets, eq.~\eqref{eq:anglechannelprofile} becomes
% }
\begin{equation}
\setlength{\abovedisplayskip}{1pt}
\setlength{\belowdisplayskip}{1pt}
    h^j[k,l]
    = \sum_{q=0}^{N_j-1}\beta_{jq} e^{-j2\pi\frac{k_{{jq}}l_{{jq}}}{NM}} \delta[k-k_{{jq}}]\delta[l-l_{{jq}}], \label{eq:AngleChannel}
\end{equation}
where $\beta_{jq}$, $k_{{jq}}$, and $l_{{jq}}$ are the coefficient, Doppler index, and delay index of the $q$-th target.
So \(h^j[k,l]\) contains weighted impulses at points $(k_{{jq}},l_{{jq}})$, $q=0,\ldots,N_j-1$.

Let us define $A'_j[k,l]$, based on the  known transmitted symbols and the estimated angles, as
{
\begin{equation}\label{eq:AmplitudePrime}
\setlength{\abovedisplayskip}{1pt}
\setlength{\belowdisplayskip}{1pt}
    A'_j[k,l]
   \buildrel \triangle \over = \sum_{n_t=0}^{N_t-1}e^{-j2\pi n_tg_t\frac{\sin(\theta_j)}{\lambda}}x_{n_t}[k,l]. 
\end{equation}
}%
% \Cref{eq:AmplitudeDecomposition} can be rewritten as
% \begin{multline}
%     A(j,l+kM)= A'(j,l+kM) \\
%     \times \sum_{q=0}^{N_j-1}\sum_{k'=0}^{N-1}\sum_{l'=0}^{M-1}h^{jq}(k-k',l-l').
% \end{multline}
% By calculating the 2D cross-correlation of $A(j, l+kM)$ and $A'(j, l+kM)$,
% Matched filtering\cref{eq:AmplitudeDecomposition} via\cref{eq:AmplitudePrime} yields
% % and omitting the amplitude of cross-correlation,
% 
% 
% % \textcolor{red}{
% For simplicity, suppose that corresponding to angle $\theta_j$ there is a single target,  with Doppler $l_0$ and delay $k_0$. 
% % \textcolor{blue}{This seems conflict with $N_j$ target in $h^j$ defined above.}
% Then it holds 
% % the channel will consist of a single peak at $[k,l]=[k_0,l_0]$ with value say $\gamma$, and in that case 
% {\small
% \begin{equation}
% \setlength{\abovedisplayskip}{1pt}
% \setlength{\belowdisplayskip}{1pt}
% A_j[k,l]= \gamma \sum_{n_t=0}^{N_t-1}e^{-j2\pi n_tg_t\frac{\sin(\theta_j)}{\lambda}} x_{n_t}[[k-k_0]_N, [l-l_0]_M],
% \end{equation}
% }%
% where $\gamma$ is the complex amplitude of $ h^j[k,l]$.
% 
% Let us define $A'_j[k,l]$, based on the  known transmitted symbols and the estimated angles, as
% {
% \begin{equation}\label{eq:AmplitudePrime}
% \setlength{\abovedisplayskip}{1pt}
% \setlength{\belowdisplayskip}{1pt}
%     A'_j[k,l]
%     =\sum_{n_t=0}^{N_t-1}e^{-j2\pi n_tg_t\frac{\sin(\theta_j)}{\lambda}}x_{n_t}[k,l]. 
% \end{equation}
% }%
From eq.~\eqref{eq:AmplitudeDecomposition} and eq.~\eqref{eq:AngleChannel}, it can be seen  that \(A_j[k,l]\) is a superposition of multiple versions of \(A'_j[k,l]\) centered at points $(k_{{jq}},l_{{jq}})$, $q=0,\ldots,N_j-1$.
Therefore, those points  can be found as the location of the peaks of the 2D cross-correlation of \(A_j[k,l]\) and \(A'_j[k,l]\), leading to target parameters 
\begin{equation}
\setlength{\abovedisplayskip}{1pt}
\setlength{\belowdisplayskip}{1pt}
\nu_{jq}=k_{{jq}}\Delta \nu={2v_{jq} f_c}/{c}, \quad \tau_{jq}=l_{{jq}}\Delta \tau={2R_{jq}}/{c}.
\end{equation}
The ability to resolve targets in the DD domain depends on both the grid spacing and the width of the autocorrelation of \(A'_j[k,l]\). 
% \textcolor{red}{there is no way to know the width of the autocorrelation. You should not make claims that you cannot prove}
% \textcolor{blue}{I see. I thought the autocorrelation in our equation is determined by the grid spacing. I was wrong about this.}
% Therefore, targets must be separated by at least twice the grid spacing to be distinguished by cross-correlation.
% This is determined by ******** the aurocorrelation will spread across multiple grid points in the DD plane. ******
% }
% \textcolor{red}{Let us define}
% % \textcolor{red}{what you propose below holds only for exclusively used bins in which case A'[] does not exist}
% \begin{align*}
%     V^j[k,l]
%     & \buildrel \triangle \over = \sum_{k'',l'',k',l'} A[j,l''+k''M]A'[j,(l''-l)+(k''-k)M]^{*}  \\
%     &= \sum_{n_t=0}^{N_t-1}\sum_{n_t'=0}^{N_t-1}e^{-j2\pi(n_t-n_t')g_t\frac{\sin(\theta_j)}{\lambda}} \sum_{k''=0}^{N-1}\sum_{l''=0}^{M-1} \sum_{k'=0}^{N-1}\sum_{l'=0}^{M-1} h^j[k',l'] \\
%     &\times  d[n_t', (l''-l')+(k''-k')M]d[n_t,(l''-l)+(k''-k)M]^{*}
% \end{align*}
% Since $d[\cdot,\cdot]$ are independent zero-mean unit power symbols (\eg QAM), the expectation of $V^j[k, l]$ is $0$ unless $k'=k$ and $l'=l$, thus
% \begin{align}
%     \E{V[k,l]} \approx N_t^2MN \cdot h^j[k,l].
% \end{align}
% % The SNR of $V^j[k,l]$ will be maximized when $k'=k$ and $l'=l$. 
% In the presence of\cref{eq:AngleChannel}, $V^j[k,l]$ will have peaks at indices $[k_{\nu_{jq}}, l_{\tau_{jq}}]$, which  can be used to estimate Range and velocity.
%
Based on the DD grid only, 
the resolution and unambiguous range/velocity are
\(R_{res}=\frac{c}{2M\Delta f} [m], \  R_{max}= \frac{c}{2\Delta f} [m], \ v_{res}=\frac{c}{2N\Delta t f_c} [m/s], \  v_{max}=\frac{c}{2\Delta t f_c} [m/s]\).
For a large unambiguous range,  \(\Delta f\) needs to be small, which results in a long symbol duration \(\Delta t\).
For a large unambiguous velocity, \(\Delta t\) needs to be small, resulting in a wide \(\Delta f\).
% This conflict is known as the \emph{Heisenberg Uncertainty Principle}.
% In real world systems, the trade-off between the range and the velocity estimation should be carefully considered.

\section{The Proposed Virtual Array for improved sensing resolution}\label{sec:VA}

In this section, we propose a novel approach to surpass the resolution limits of the radar receive array by creating a virtual array and making selective trade-offs between sensing performance and communication rate.
To achieve this, we introduce the concept of \textit{TF domain private bins}, each exclusively paired with a specific transmit antenna.

\subsection{Radar sensing}
Let us, for simplicity, first present the idea for the case of {$N_t=2$ transmit antennas.}
Let bin $[n_p,m_p]$  be exclusively paired with transmit antenna $p$ (or otherwise, private to antenna $p$) for $p=1,2$, and let 
$X_{p}[n,m]$ represent the  time-frequency (TF) signal of the \(p\)-th antenna.
Since bin $[n_2,m_2]$ is for  use by antenna $2$ only, antenna $1$ 
sets $X_{1}[n_2, m_2]=0$. Similarly, since bin $[n_1,m_1]$ is for the use of antenna $1$ only, antenna $2$ sets  $X_{2}[n_1, m_1]=0$.

\pgfdeclareverticalshading{rainbow}{80bp}{
    color(0bp)=(red!25); color(30bp)=(red!25); color(50bp)=(blue!25); color(80bp)=(blue!25)
    }
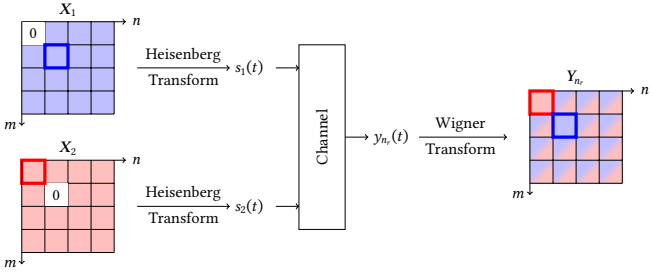
\begin{figure}
    \centering
    \resizebox{\linewidth}{!}{%
    \begin{tikzpicture}[every node/.style={minimum size=.5cm-\pgflinewidth, outer sep=0pt},shading=rainbow]
    
        \node[above] at (5,1) {${\mathbf{X}}_{1}$};
        \fill[blue!25] (4,1) rectangle (6,-1);
        \draw[step=0.5cm,color=black] (4,1) grid (6,-1);
        \node[fill=white] at (4.25,0.75) {0};
        \draw[color=blue,line width=2pt] (4.5,0.5) rectangle (5,0);
        \draw[->] (6,1) -- (6.25,1) node[right] {$n$};
        \draw[->] (4,1) -- (4,-1.25) node[left] {$m$};    
        \draw[->] (6.5,0) -- (8.5,0);
        \node[above] at (7.5,0) {Heisenberg};
        \node[below] at (7.5,0) {Transform};    
        \node[right] at (8.5,0) {$s_1(t)$};
        \draw[->] (9.5,0) -- (10,0);

        \node[above] at (5,-2) {${\mathbf{X}}_{2}$};
        \fill[red!25] (4,-2) rectangle (6,-4);
        \draw[step=0.5cm,color=black] (4,-2) grid (6,-4);
        \node[fill=white] at (4.75,-2.75) {0};
        \draw[color=red,line width=2pt] (4,-2) rectangle (4.5,-2.5);
        \draw[->] (6,-2) -- (6.25,-2) node[right] {$n$};
        \draw[->] (4,-2) -- (4,-4.25) node[left] {$m$};    
        \draw[->] (6.5,-3) -- (8.5,-3);
        \node[above] at (7.5,-3) {Heisenberg};
        \node[below] at (7.5,-3) {Transform};    
        \node[right] at (8.5,-3) {$s_2(t)$};
        \draw[->] (9.5,-3) -- (10,-3);

        \draw[black] (10,0.5) rectangle (11,-3.5);
        \node[rotate=90] at (10.5, -1.5) {Channel};
        \draw[->] (11,-1.5) -- (11.5,-1.5);
        \node[right] at (11.5,-1.5) {$y_{n_r}(t)$};
        \draw[->] (12.6,-1.5) -- (14.5,-1.5);
        \node[above] at (13.5,-1.5) {Wigner};
        \node[below] at (13.5,-1.5) {Transform};    

        \node[above] at (16,-0.5) {$\mathbf{Y}_{n_r}$};
        \foreach \x in {16.5, 15, 15.5, 16}
        \shade[shading angle=45] (\x,-0.5) rectangle +(0.5,-0.5);
        \foreach \x in {16.5, 15, 15.5, 16}
        \shade[shading angle=45] (\x,-1) rectangle +(0.5,-0.5);
        \foreach \x in {16.5, 15, 15.5, 16}
        \shade[shading angle=45] (\x,-1.5) rectangle +(0.5,-0.5);
        \foreach \x in {16.5, 15, 15.5, 16}
        \shade[shading angle=45] (\x,-2) rectangle +(0.5,-0.5);
        \fill[red!25] (15,-0.5) rectangle (15.5,-1);
        \fill[blue!25] (15.5,-1) rectangle (16,-1.5);
        \draw[step=0.5cm,color=black] (15,-0.5) grid (17,-2.5);
        \draw[color=red,line width=2pt] (15,-0.5) rectangle (15.5,-1);
        \draw[color=blue,line width=2pt] (15.5,-1) rectangle (16,-1.5);
        \draw[->] (17,-0.5) -- (17.25,-0.5) node[right] {$n$};
        \draw[->] (15,-1) -- (15,-2.75) node[left] {$m$};    
        
    \end{tikzpicture} 
    }%
    \caption{Orthogonal  received signals on private TF bins.}
    \label{fig:VA}
\end{figure}

The TF signal received by radar receive antenna $n_r$ on  private bin $[n_p,m_p]$ equals (see eq.~\eqref{eq:TFIO},~\eqref{eq:TFChannel}, and fig.~\ref{fig:VA})
\begin{multline}
\setlength{\abovedisplayskip}{0pt}
\setlength{\belowdisplayskip}{0pt}
    Y_{n_r}[n_p, m_p]
    =\sum_{j=0}^{J_T-1} e^{-j2\pi(n_r g_r + p g_t)\frac{sin(\theta_j)}{\lambda}}  X_{p}[n_p,m_p] \\
    \times \beta_j e^{-j2\pi \nu_j \tau_j}e^{j2 \pi(\nu_j n_p \Delta t - m_p \Delta f \tau_j)}. \quad p=1,2\label{16}
\end{multline}    
% for $p=1,2$. 
% The radar receive antenna $n_r$, knowing the transmitted symbols, can formulate the ratio
% \begin{equation}
% R^{n_r}[n_l,m_l] \buildrel \triangle \over=  \frac{Y_{TF}^{n_r}[n_l,m_l]}{X_{TF}^{l}[n_l,m_l]}.
%     \end{equation}
Placing the ratio \( {Y_{n_r}[n_p,m_p]}/{X_{p}[n_p, m_p]}\) of all receive antennas in vector $\mathbf{r}_p$, we get
\begin{equation}
\setlength{\abovedisplayskip}{1pt}
\setlength{\belowdisplayskip}{1pt}
    \mathbf{r}_p=\sum_{j=0}^{J_T-1}\beta_j 
    \mathbf{a}_r(\theta_j, \nu_j,\tau_j ; n_p,m_p). \label{17}
\end{equation}
\(\mathbf{r}_p \)  can be viewed as the output of a linear array of size $N_r$. By stacking \(\mathbf{r}_1 \) and \(\mathbf{r}_2 \) into vector \(\mathbf{r} \), we can formulate an effective virtual array of size $2N_r$.
Further, we express \(\mathbf{r} \) as
\begin{equation}
    \mathbf{r}=[\mathbf{r}^T_1 \ \mathbf{r}^T_2]^T = \mathbf{\Phi}\bm{\beta},
\end{equation}
where $\mathbf{\Phi}$ is an overcomplete matrix, whose columns are
\([\mathbf{a}^T_r(\tilde \theta_j, \tilde \nu_j,\tilde\tau_j ; n_1,m_1) \ \mathbf{a}^T_r(\tilde \theta_j, \tilde \nu_j,\tilde\tau_j ; n_2,m_2)]^T\), with (\(\tilde \theta_j, \tilde \nu_j,\tilde\tau_j\))
corresponding to a grid point of the discretized angle-velocity-delay space.
Each element of the vector \(\bm{\beta}\) corresponds to a grid point in the target space, where a non-zero value indicates the presence of a target at that location, and a zero value indicates no target.
Since we already have preliminary target parameters estimated as described in sec.~\ref{sec:target}, we can discretize the target space around those estimates. 
By restricting the target space to a specific angular region, 
\( \bm{\beta} \) becomes a sparse vector, with only a few non-zero elements corresponding to the actual target locations.
Therefore, we can solve the following sparse signal recovery (SSR) problem~\cite{6650099} to estimate \( \bm{\beta} \)
\begin{equation}\label{eq:SSR}
    \min_{\mathbf{\beta}}  \norm{\mathbf{r}-\mathbf{\Phi}\bm{\beta}}_2^2+\lambda\norm{\bm{\beta}}_1.
\end{equation}
{This design can be generalized to $N_p$ private bins and $N_p$ is up to $N_t$.}
% \textcolor{red}{Can you check  whether  this enables you to achieve better resolution than the one allowed by your choice of $M,N$??
% For example, simulate the targets based on $N^*,M^*$ larger than the $N,M$  that you use to implement the OTFS. Where will the targets appear during the coarse angle estimation? Hopefully they will be approximated by targets on  the $N,M$ grid. Can the sparse signal recovery recover the targets closer to the $N^*,M^*$ grid?
 % }
% \textcolor{blue}{We refer to the TF bins used exclusively as private bins, and the vector $\mathbf{r}$ can be viewed as virtual array.}
% The detailed procedure will be explained in~\cref{sec:VA_sim}.

\subsection{Communication}
While modifying (setting to zero)  TF domain bins leads to diversity that can be exploited to improve target resolution, it does not allow going back to the DD domain via an SFFT. Therefore, it has an effect on the set of communication symbols that need to be sent to the communication receiver.
To avoid this effect, we propose to restructure the DD domain.

Let each antenna include a  zero among its DD-domain symbols. By doing so, each antenna decreases by $1$ its information bearing symbols, which are now $NM-1$.
Each ISFFT point,  $X_{p}[n,m]$, is a linear combination of all the DD domain symbols of the $p$th antenna. 
Even if antenna
$p$ destroys one TF bin by setting it to zero,  the remaining 
$NM-1$ bins provide 
$NM-1$ linear combinations of the antenna's 
$NM-1$  DD domain information-bearing symbols.

% For example, suppose that antenna 
% $p$  places a $0$ in place of $x_{p}[0,0]$, and  then  sets $X_{p}[0,0]=0$. 
The ISFFT of $x_p[k,l]$  can be represented in vector form as
\begin{equation}
\mathbf{X}_{p} = \mathbf{G}\mathbf{x}_{p}.
\end{equation}
\(\mathbf{X}_{p}\) has in its $(m+nM)$-th position $X_p[n,m]$; \(\mathbf{x}_{p}\) has in its $(l+kM)$-th position \(x_p(k,l)\);
 $\mathbf{G}=(\mathbf{F}_N^H\otimes\mathbf{F}_M)$, with 
$\mathbf{F}_M$ being an  $M$-point DFT matrix; and 
$\otimes$ denoting  Kronecker product.

% Suppose that antenna $1$ sets  $X_1[n_2,m_2]=0$. 
{
Continuing with the $N_T=2$ antenna case,
}%
we define vector \(\tilde {\mathbf{X}}_{1}\) to be constructed from \(\mathbf{X}_{1}\) by excluding $X_1[n_2,m_2]$ (\ie., containing the RF symbols shown in blue color in fig. \ref{fig:VA}).
As long as  antenna $1$ had set  $x_1[k_0,l_0]=0$, for some $k_0,l_0$,  its non-zero DD bins can be obtained from \(\tilde {\mathbf{X}}_{1}\) 
 as
\begin{equation}
 \tilde{\mathbf{x}}_{1}=  \mathbf{C}_{1}\inv \tilde{\mathbf{X}}_{1}. \label{21}
\end{equation}
In the above, \(\tilde{\mathbf{x}}_{1}\) is constructed from \(\mathbf{x}_{1}\) by excluding $x_1[k_0,l_0]$, and 
$\mathbf{C}_1$ is constructed based on  $\mathbf{G}$, by removing its $(m_2+n_2M)$-th  row and its $(l_0+k_0M)$-th column.

In the more general case, where private TF bins are assigned to $N_p$ (out of the $N_t$) transmit antennas, $N_p-1$ zeros need to be inserted in each transmit antenna's TF signal.
To ensure recovery of the antenna's information-bearing symbols, $N_p-1$ DD domain symbols need to be set to zero for each transmit antenna as well.
Thus, in order to refine the sensing resolution by $N_p$, the total number of information bearing symbols are reduced to 
$N_t N M - N_p (N_t-1)$. 
As will be shown in the simulations section,  
significant sensing benefit can be achieved with a small $N_p$.

% \textcolor{red}{verify this with simulations}.
% \textcolor{blue}{I found with SSR, we can achieve high resolution with just $2$ private bins. More private bins will greatly speed up the computation.}

% \noindent\textbf{Communication}%

% Assume that all bins are available at all antennas. The $N_t$ transmit antennas send signals to $N_c$ receivers antennas for communication purpose. 
% \vspace{-0.2in}
Assuming that $N_c\geq N_t$, and after performing demodulation, 
% On vectorizing \cref{eq:DemodulatedSymbol}
% }
the  DD domain I/O across all receive antennas can be expressed as in~eq.\eqref{eq:vector_io}.
{
% \small
\begin{equation*}\label{eq:MIMO}
    \underbrace{
    \begin{bmatrix}
        \mathbf{y}_{1} \\
        \vdots \\
        \mathbf{y}_{N_c}
    \end{bmatrix}
    }_{\mathbf{y}_{MIMO}\in\C^{N_cNM}}
    =
    \underbrace{
    \begin{bmatrix}
        \mathbf{h}_{(1,1)} & \cdots & \mathbf{h}_{(1,N_t)} \\
        \vdots & \ddots & \vdots \\
        \mathbf{h}_{(N_c,1)} & \cdots & \mathbf{h}_{(N_c,N_t)}
    \end{bmatrix}
    }_{\mathbf{h}_{MIMO}\in\C^{N_cNM\times N_tNM}}
    \underbrace{
    \begin{bmatrix}
        \mathbf{x}_{1} \\
        \vdots \\
        \mathbf{x}_{N_t}
    \end{bmatrix}
    }_{\mathbf{x}_{MIMO}\in\C^{N_tNM}}
    +
    \underbrace{
    \begin{bmatrix}
        \mathbf{w}_{1} \\
        \vdots \\
        \mathbf{w}_{N_c}
    \end{bmatrix}
    }_{\mathbf{w}_{MIMO}\in\C^{N_cNM}},
\end{equation*}
}%
% \begin{equation}\label{eq:vector_MIMO}
% \hat{\mathbf{d}}_{MIMO}=\mathbf{H}_{MIMO}\mathbf{d}_{MIMO}+\mathbf{w}_{MIMO}.
% \end{equation}
where $\mathbf{h}_{MIMO}$ is the MIMO DD effective channel matrix with $N_c\times N_t$ blocks and each block is the DD effective channel matrix $\mathbf{h}_{(n_c, n_t)}$ between the $n_c$ receive antenna and the $n_t$ transmit antenna as defined in eq.~\eqref{eq:vector_io}.
On assuming every $\mathbf{h}_{(n_c, n_t)}$ has been estimated via a pilot,
% \textcolor{red}{define the elements of this matrix - we have multipath here}
% \textcolor{blue}{Added}
% $\mathbf{d}_{MIMO}\in\C^{N_t NM}$
% =\mathrm{vec}\braces*{\mathbf{D}\trn}$ 
% is the vectorized transmitted symbol matrix $\mathbf{D}$, whose element on the $n_t$ row and $l+kM$ column is $d[n_t, l+kM]$.
% constructed as follows: its $n_t$-th row  contains the symbols transmitted by the $n_t$-th antenna on all bins, and its 
%  $l+kM$-th column  contains the symbols transmitted by all antennas on bin \([k,l]\).
% $\mathbf{w}_{MIMO}\in\C^{N_tNM}$ is the vector of white noise.
% with noise power $\sigma_{\mathbf{w}}^2$.
 % and $\mathrm{vec}\braces*{\cdot}$ is the vectorization operator.
% $\hat{\mathbf{d}}_{MIMO}\in\C^{N_c NM}$ is the vectorized received symbol matrix from the same vectorization operation.
the symbol vector 
% \textcolor{red}{bad symbol; reminds of derivative. use $\hat x$}
$\hat{\mathbf{x}}_{n_t}$ for all $n_t$, can be estimated via LMMSE equalization.
% via Least Squares, \(\argmin_{\mathbf{d}\in\mathbb{A}^{N_tNM}}\norm*{\hat{\mathbf{d}}-\mathbf{H}\mathbf{d}}_2^2\), where $\mathbb{A}^{N_tNM}$ is the symbol constellation set with $N_tNM$ possible elements.
% By applying the same process to every bin and every received OTFS signal, the communication receiver can estimate all transmitted symbols.
% \textcolor{blue}{
% $\dot{\mathbf{d}}_{MIMO}$ has $N_t$ segments with length $NM$, and each segment $\dot{\mathbf{d}}^p$ is the estimated symbol vector of $p$-th transmit antenna.
% The communication receivers then calculates the TF domain equivalent of $\dot{\mathbf{d}}^{p}$, \ie $\dot{\mathbf{X}}_{TF}^p$, via ISFFT.
% With the knowledge of the position of zeros in the TF domains on each transmit antenna, the zeros introduced in each antenna at the transmitter can be enforced and the symbol vectors become $\tilde{\mathbf{X}}_{TF}^p$.
% Based on knowledge of the structure of the signal in the DD domain, $\mathbf{C} ^{p}$ can be constructed at the receiver and the information bearing symbols can then be recovered along the lines of \cref{21}.
% }

Based on $\hat{\mathbf{x}}_{n_t}$, the information bearing symbols can be obtained by taking an ISFFT to get to the TF domain, extracting the nonzero TF bins, and then applying~eq.\eqref{21}. This assumes knowledge of where the zero bins in the DD and TF domains are.
Compared to the MIMO communication system without shared bins, the proposed system increases the number of bits transmitted in one period by at most a factor of $N_t$.
The communication rate grows linearly with the grid size, \(N\) and \(M\), which also enhances sensing resolution. However, larger grid sizes increase equalization complexity, so careful selection of system parameters is essential in DFRC MIMO OTFS system design.

% \textcolor{magenta}{so, we can keep the grip size small, to reduce comm equalization complexity, and still get good sensing resolution via the virtual array.}

% \textcolor{red}{need some results on how closely you can resolve targets in angle and DD place, and how that min distance depends on various parameters, N,M, Nr, Nt}

% \section{Trading off Communication Rate for Angle Estimation Using Shared-Private Design}\label{sec:tradeoff}
% \input{03tradeoff.tex}

% \vspace{-0.4in}
\section{Simulation Results}\label{sec:simulation}
% \vspace{-0.2in}
{
% \tiny
\begin{table}[H]
    \centering
    \caption{System Parameters}
    \label{tab:sys_parameters}
    % \resizebox{\columnwidth}{!}{%
    \begin{tabular}{|c|c|c|}
    \hline
    Symbol      & Parameter             & Value         \\ \hline \hline
    \(M\)        & Number of subcarriers & $128$         \\ \hline
    \(N\)        & Number of subsymbols  & $64$          \\ \hline
    % \(B\)        & Total bandwidth       & $10$ MHz      \\ \hline
    \(\Delta f\) & Subcarrier spacing    & $120$ KHz  \\ \hline
    \(f_c\)      & Carrier frequency     & $24.25$ GHz      \\ \hline
    % \(R_{res}\)  & Range resolution      & $15$ m        \\ \hline
    % \(V_{res}\)  & Velocity resolution   & $3.8125$ m/s  \\ \hline
    % \(R_{max}\)  & Unambiguous range     & $3840$ m      \\ \hline
    % \(V_{max}\)  & Unambiguous velocity  & $\pm 122$ m/s \\ \hline
    \(g_t\)      & Transmit antenna spacing & $0.5\lambda$ \\ \hline
    \(g_r\)      & Receive antenna spacing  & $0.5\lambda$ \\ \hline
    % \(J\)        & Number of targets/paths     & $3$           \\ \hline
    % \(\theta_j\) & Angle of targets/paths \  & $[7, 25, -25]^{\circ}$ \\ \hline
    % \(R_j\)      & Range of targets/paths   & $[194.87, 104.93, 149.90]$ m \\ \hline
    % \(v_j\)      & Velocity of targets/paths  & $[-89.98, 76.13, -117.66]$ m/s \\ \hline
    \end{tabular}%
     % }
\end{table}
}%

In this section, we present simulation results to demonstrate the performance of the proposed virtual array.
The simulated system parameters follow the 5G NR high-frequency standard~\cite{5gnr} and are shown in table~\ref{tab:sys_parameters}.
The high-Doppler channel is simulated by the Matlab function
~\cite{BibEntry2024Jun} 
with targets/paths parameters shown in table~\ref{tab:share_parameters},~\ref{tab:private_parameters}.
% Compared to~\cite{10036975}, where the targets' speed is at most $21.12 [m/s]$, we here consider high-speed targets, which will cause high Doppler shifts and will make OFDM fail. 
% \textcolor{red}{is this both for sesning and communication??}
% \textcolor{blue}{
% The channels for both are simulated by the same function except the sensing channel contains steering vectors. 
% The sensing channel *****describe and give equation number *******. The communication channel contains **** multipaths.
% }
The sensing channel contains $3$ targets.
The communication channel contains $3$ paths, and different pairs of transmit and communication antennas have different complex path gains. 
The information symbols are QPSK, the signal-to-noise ratio (SNR) is $20$ dB, and the guard interval is a cyclic prefix with the length equal to the maximum delay introduced by the targets.

{
% \tiny
\begin{table}[H]
    \centering
    \caption{Channel Parameters}
    \label{tab:share_parameters}
    % \resizebox{\columnwidth}{!}{%
    \begin{tabular}{|c|c|c|}
    \hline
    Symbol      & Parameter             & Value         \\ \hline \hline
    \(J\)        & Number of targets/paths     & $3$           \\ \hline
    \(\theta_j\) & Angle of targets/paths \  & $[-25, 7, 15]^{\circ}$ \\ \hline
    \(R_j\)      & Range of targets/paths   & $[68.31, 78.07, 48.79]$ m \\ \hline
    \(v_j\)      & Velocity of targets/paths  & $[57.95, -104.31, 81.13]$ m/s \\ \hline
    \end{tabular}%
     % }
\end{table}
}%

\begin{figure}%[!htbp]
    \centering
    \begin{subfigure}[t]{0.5\columnwidth}
        \centering
        \includegraphics[width=\columnwidth]{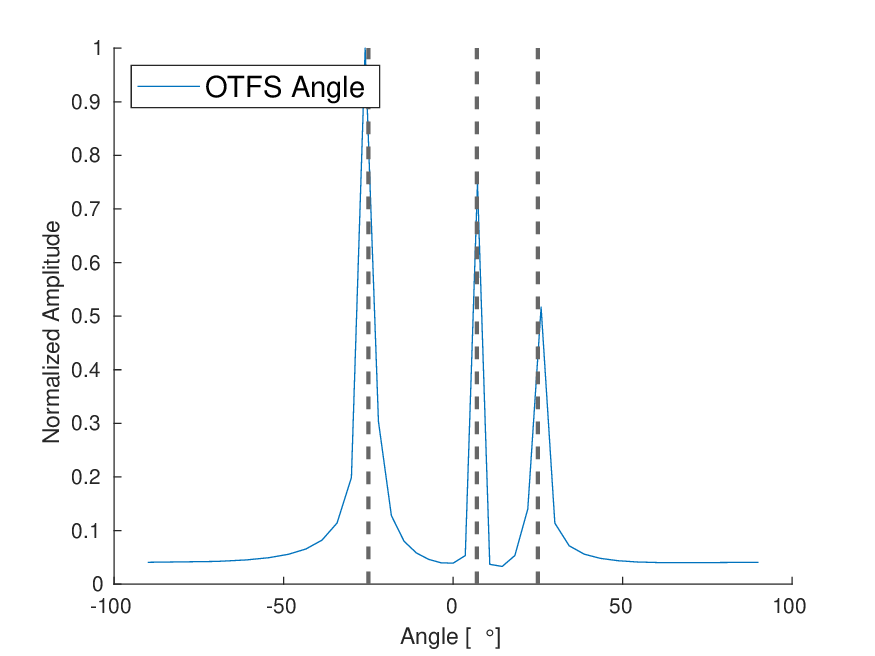}
        \caption{}
        \label{fig:angle_8x16}
    \end{subfigure}%
    ~
    \begin{subfigure}[t]{0.5\columnwidth}
        \centering
        \includegraphics[width=\columnwidth]{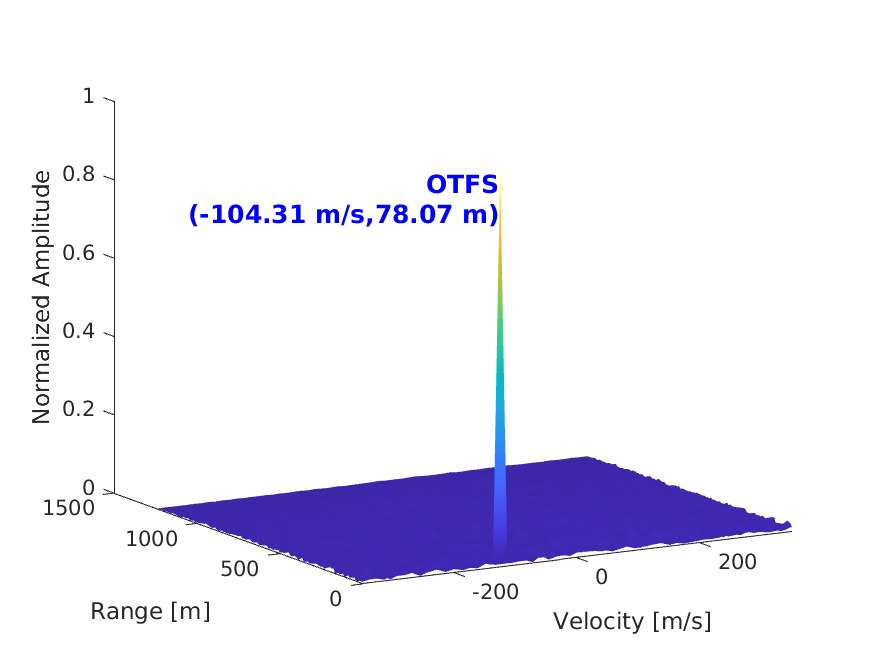}
        \caption{}
        \label{fig:2D_128x32}
    \end{subfigure}
    % \begin{subfigure}[t]{0.5\columnwidth}
    %     \centering
    %     \includegraphics[width=\columnwidth]{fig/range.eps}
    %     \caption{}
    %     \label{fig:2D_OTFS_128x32}
    % \end{subfigure}%
    % ~
    % \begin{subfigure}[t]{0.5\columnwidth}
    %     \centering
    %     \includegraphics[width=\columnwidth]{fig/Velocity.eps}
    %     \caption{}
    %     \label{fig:2D_OFDM_128x32}
    % \end{subfigure}
    \caption{Radar performance of the OTFS system with all DD bins used as shared.
    % and comparison to the OFDM system of ~\cite{10036975} under the same setup. 
    Fig.~\ref{fig:angle_8x16} shows angle estimation with ground truth indicated by black dash lines. Fig.~\ref{fig:2D_128x32} shows the 2D Cross-correlation corresponding to the second target. 
    % \Cref{fig:2D_OTFS_128x32,fig:2D_OFDM_128x32} are~\cref{fig:2D_128x32} along range and velocity axes, respectively.
    }
    \label{fig:radar}
\end{figure}

\subsection{Sensing with all DD bins used as shared}

% \smallskip
The targets/paths parameters are as shown in table~\ref{tab:share_parameters},  $N_t=4$, $N_r=32$. 
In this case, 
all targets are well separated.
The target angles were estimated via the DFT approach described in sec.~\ref{sec:target}, and the result is shown in fig.~\ref{fig:angle_8x16}.
% Both OTFS and OFDM systems successfully estimate the correct angles of $3$ targets in the channel as shown in~\cref{fig:angle_8x16}. which has sufficient angle resolution from adequate number of receive antennas $N_r=16$.
% When the system only has $N_r=8$ receive antennas, both OTFS and OFDM fail to reveal all targets as shown in~\cref{fig:angle_8x8}.
The range and velocity of the $2$nd target was estimated via the 2D cross-correlation approach of sec.~\ref{sec:target} and is shown in fig.~\ref{fig:2D_128x32}.

\subsection{Trading off Communication for Sensing Using Private Bins}\label{sec:VA_sim}
\begin{figure}%[!htbp]
    \centering
    \begin{subfigure}[t]{0.5\columnwidth}
        \centering
        \includegraphics[width=\columnwidth]{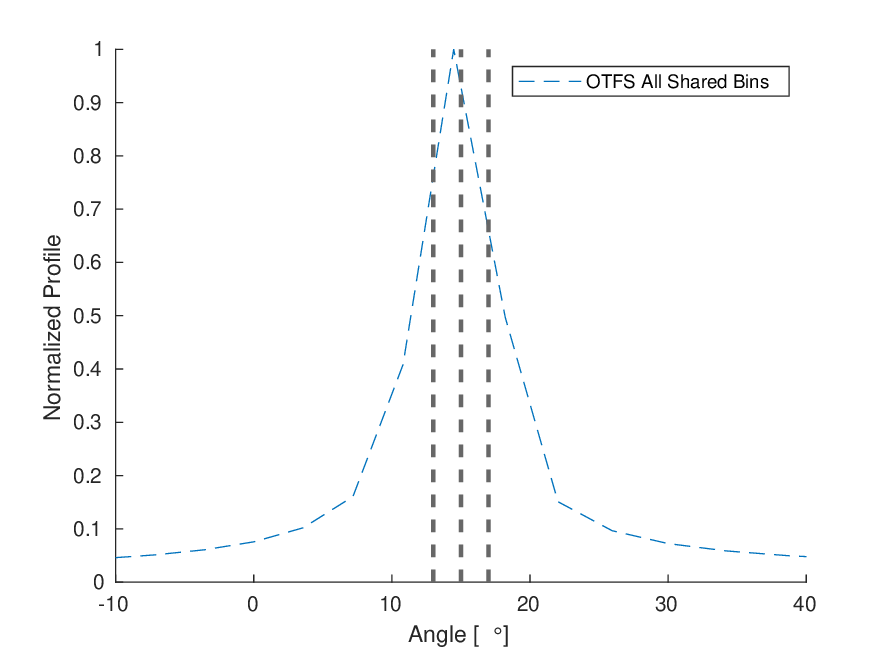}
        \caption{}
        \label{fig:OTFS_VA_DFT}
    \end{subfigure}%
    ~
    \begin{subfigure}[t]{0.5\columnwidth}
        \centering
        \includegraphics[width=\columnwidth]{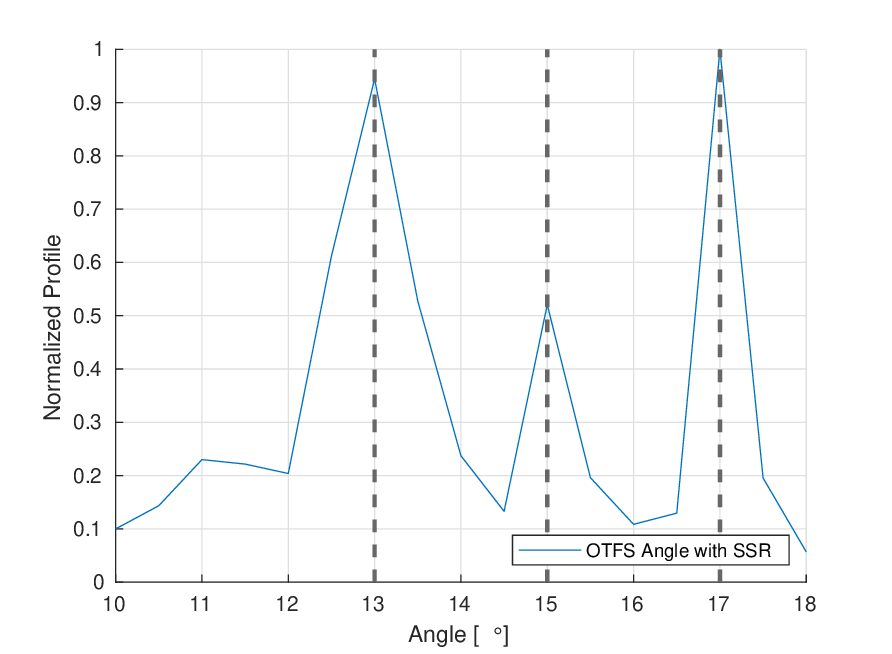}
        \caption{}
        \label{fig:OTFS_VA_SSR}
    \end{subfigure}
    \caption{Resolution of DFT- and SSR-based angle estimation.}
    \label{fig:close_angle}
\end{figure}

{
% \tiny
\begin{table}%[H]
    \centering
    \caption{Channel Parameters}
    \label{tab:private_parameters}
    % \resizebox{\columnwidth}{!}{%
    \begin{tabular}{|c|c|c|}
    \hline
    Symbol      & Parameter             & Value         \\ \hline \hline
    \(J\)        & Number of targets/paths     & $3$           \\ \hline
    \(\theta_j\) & Angle of targets/paths \  & $[17, 13, 15]^{\circ}$ \\ \hline
    \(R_j\)      & Range of targets/paths   & $[68.31, 48.79, 78.07]$ m \\ \hline
    \(v_j\)      & Velocity of targets/paths  & $[46.36, -139.08, 81.13]$ m/s \\ \hline
    \end{tabular}%
     % }
\end{table}
}%
% \textcolor{red}{Explain what you did, connect it to the appropriate section. You cannot start from the end. Rewrite the text below. There has to be a logical sequence. }
% Next, we use the received TF domain signals on private bins to refine the radar parameter estimation. The channel parameters are shown in~\cref{tab:private_parameters} with $3$ targets located in closely spaced angle bins. For the same radar system setup (\cref{tab:sys_parameters}), we consider $N_p=4$ TF bins to be private, and the loss of communication rate is $0.37\%$.
We take $N_t=4$, $N_r=32$. When the targets are closely spaced in angle, their resolvability depends on the resolution of the receive array, which for $N_r=32$ is $\pi/N_r=5.625^{\circ}$.
Here, the channel parameters are shown in table~\ref{tab:private_parameters}. With three targets located within $2^{\circ}$, only one peak is visible in the DFT, as shown in fig.~\ref{fig:OTFS_VA_DFT}.
Next, we use the private bins designed as described in sec.~\ref{sec:VA},to formulate and solve an SSR problem to refine  angle estimation.

% \textcolor{red}{First we show how target estimation works where all bins are used as shared. ***** run simulations for a case where the targets are resolvable. Show results.}
% \textcolor{blue}{This is in the previous subsection.}
% \medskip
% \noindent{\textit{Formulating a virtual array}}
% \textcolor{red}{
% When targets appear closer in angle, their resolvability depends on the aperture of the receive array.
% In our example, 
% this means that when the targets are closer than $\frac{\pi}{(N_r-1)g_r/\lambda}^{\circ}$, they cannot be resolved. 
% We illustrate for the case of targets' angle spacing as close as $2^{\circ}$, where they cannot be resolved.
% As shown in~\cref{fig:OTFS_VA_DFT},  a single peak is visible at the center of the three targets.
%
% \Cref{fig:close_angle} presents the results of angle estimation. 
We used $4$ private bins. 
Specifically, for each transmit antenna, we set the DD bins $[0,0]$, $[1,1]$, and $[2,2]$,
to zero and placed symbols on the remaining DD grid. 
After performing the ISFFT, we obtained the TF domain signal and introduced TF private bins at positions $[0,0]$, $[1,1]$, $[2,2]$ and $[3,3]$.
Each transmit antenna was assigned one TF private bin, while the remaining $3$ bins were set to zero.
% Based on the received signal corresponding to private TF bins, we construct a sparse signal recovery problem. 
% \textcolor{red}{delete the red line in eq. 6. I am not sure that you are formulating a VA correctly unless you show me that by increasing the number of private bins the resolution improves} 
% \textcolor{blue}{
% The VA we constructed is a sequence with length P (number of private bins) times of eq 14. 
% This does not change aperture, which is restricted by $N_r$.
% But this longer sequence will have a finer spectrum, so we will be able to see the number of targets. 
% We will need the number of targets to determine when the iterative detection stops.
% }
% While the virtual array enables the detection of three peaks using the DFT detector, its aperture remains limited by the number of receive antennas $N_r$ and the number of private bins $P$, i.e., $(N_rP - 1)g_r$. Consequently, the detected peaks do not align with the true target positions, as indicated by the black dashed line.
% }%
% To achieve higher resolution, we formulated an SSR (sparse signal recovery) problem. 
We then formulated the SSR problem with the help of the signal received on the private TF bins.
% The angle space around the coarse estimated angles was discretized, paired with the range and velocity estimated by 2D cross-correlation, and the base matrix $\Phi$ was constructed as described in~\cref{sec:VA}. 
% Each column of $\Phi$ corresponds to a discretized angle along with the estimated ranges and velocities.
Upon solving the SSR problem, the peaks align accurately with the ground truth, as demonstrated in fig.~\ref{fig:OTFS_VA_SSR}.
In our simulations, we set  $\lambda=10^{-5}$, which  balances resolution and computation effectively.

% \begin{figure}
%     \centering
%     \includegraphics[width=0.5\linewidth]{fig/DFT_angle_close_target.eps}
%     \caption{2D cross-correlation with coarse angle estimated}
%     \label{fig:DFT_angle_close_target}
% \end{figure}

% To ensure the resolvability of the SSR detector, we need to address three key questions:
% \begin{enumerate*}
%     \item How do we confirm all targets within the angle space are detected?
%     \item How do we choose the optimal penalty parameter $\lambda$ in~\cref{eq:SSR}?
%     \item How fine should the angle space be discretized?
% \end{enumerate*}
% The first question can be addressed through 2D cross-correlation range-velocity estimation using coarse DFT angle. 
% As explained in~\cref{sec:target}, when multiple targets share the same angle profile, the radar detector shows multiple peaks.
% We apply SSR until reveal angles of all peaks. \textcolor{red}{more information is needed here. You follow a deflation-type approach....}
% % The second and third questions affect SSR's computational complexity, which depends on the penalty parameter $\lambda$ and the size of matrix $\Phi$. 
% % A smaller $\lambda$ offers higher resolution but slower convergence, while a larger $\lambda$ converges faster but may miss targets. 

% The choice of the penalty parameter $\lambda$ in~\cref{eq:SSR} plays a big role in solving the SSR problem.
% Typically, $\lambda$ is chosen empirically~\cite{Hastie2020}. 
% In our simulations, $\lambda=10^{-5}$ balances resolution and computation effectively.

The discretization of the target space is a critical factor in resolving multiple targets. If the discretization is too coarse, multiple targets may still fall into the same bin. To address this, we propose an iterative refinement mechanism. In our simulations, we initially set the angular spacing to \(d_{\alpha} = \floor{\pi/2 N_r}^\circ \). We then run the SSR detector and iteratively remove the peak with the highest magnitude. If a peak reappears at the same angular bin but corresponds to different Doppler and delay bins during the iteration, it indicates that the grid spacing is too coarse. In such cases, we refine the spacing to \(d_{\alpha+1} = d_{\alpha}/2\)  and rerun the detector. The process continues until all targets are resolved. 
% This approach is detailed in  ~\cref{alg:SSR}.

\begin{figure}
\centering
\begin{subfigure}[t]{0.7\columnwidth}
    \centering
    \includegraphics[width=\columnwidth]{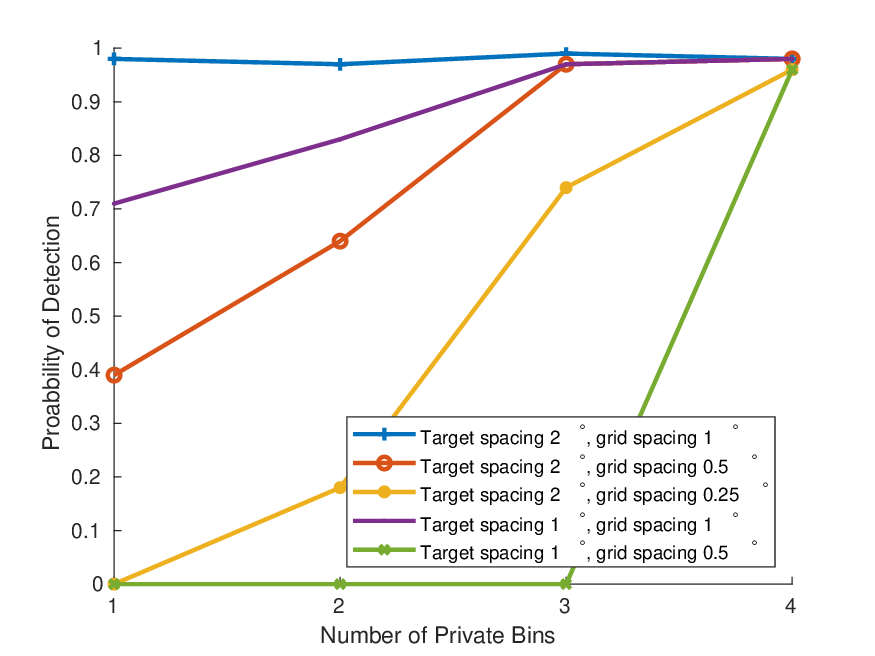}
    % \caption{SSR Angle Detection Capability}
    % \label{fig:SSR}
\end{subfigure}%
% ~
% \begin{subfigure}[t]{0.5\columnwidth}
%     \centering
%     % \includegraphics[width=\linewidth]{fig/OTFS_private.eps}
%     \includegraphics[width=\columnwidth]{fig/OTFS_private_2.eps}
%     \caption{Communication BER. $N_c=8$.}
%     \label{fig:private_communication}
% \end{subfigure}
% \caption{DFRC MIMO system performance using private bins.}
\caption{SSR Angle Detection Capability.}
\label{fig:SSR}
\end{figure}

Next, we illustrate how the number of private bins and angle space discretization can affect the detection probability of the SSR approach. We performed $100$ Monter Carlo simulations; 
in each simulation $3$ targets, separated in angle by at least $2^{\circ}$ were randomly placed in the angle space  $[-60, 60]^{\circ}$.
% and guarantee their spacing is $2^{\circ}$ or $1^{\circ}$.
% For the experiment of targets with $2^{\circ}$ spacing, we consider 
The experiment was repeated for  different number of private bins 
% was $[1, 0.5, 0.25]^{\circ}$, and for the experiment of targets with $1^{\circ}$ spacing, we consider initial discretization spacing $[1, 0.5]^{\circ}$.
and different levels of angle space discretization.
The probability of detection versus number of private bins  is shown in fig.~\ref{fig:SSR}.
The experiment was repeated for 
$3$ target separated in angle by at least $1^{\circ}$ and the result is also shown in fig.~\ref{fig:SSR}. We can observe that  increasing the number of private bins  greatly improves the probability of detection, as it  increases the number of rows in $\mathbf{\Phi}$.
% The more private bins, the more diversity can be used to solve the SSR problem.
Also, finer discretization does not always lead to better detection, as it may increase the coherence of $\mathbf{\Phi}$.
Finally, we observe that when the targets  minimal angle separation is $2^{\circ}$, it is possible to achieve perfect detection with only $1$ private bin.
% \textcolor{red}{read the above paragraph. Is it ok?}
% \textcolor{blue}{looks good.}

% \begin{algorithm}
% \caption{SSR angle detection}\label{alg:SSR}
% \begin{algorithmic}[1]
% \Require $\mathbf{r}$, $\mathbf{\beta}$, $\theta_{j}$, $\nu_{jq}\in\mathbf{\nu}_j$, $\tau_{jq}\in\mathbf{\tau}_j$, $n_p$, $m_p$, $N_r$, $N_j$, $\lambda=10^{-5}$
% \Ensure $N_j$ number of $\theta_{jq}$
% \State $\alpha=1$, $d_{\alpha}=\floor{\pi/2N_r}^{\circ}$
% \State $\tilde{\theta}=(\theta_j-\alpha d_{\alpha}):d_{\alpha}:(\theta_j+\alpha d_{\alpha})$ \label{step2}
% \State Construct $\Phi$ whose columns are $\mathbf{a}_r^T(\tilde{\theta},\nu_{jq},\tau_{jq};n_p,m_p)$ 
% \While{Number of unique $\theta_{jq} \neq N_j$}
% \State $\hat{\mathbf{\beta}}_{jq}=\argmin  \norm{\mathbf{r}-\mathbf{\Phi}\bm{\beta}}_2^2+\lambda\norm{\bm{\beta}}_1$
% \State Set $\theta_{jq}$ to the index of the largest magnitude in $\hat{\bm{\beta}}_{jq}$
% \State Find corresponding $\nu_{jq}$, $\tau_{jq}$ 
% \State Update $\mathbf{r}$ and $\mathbf{\Phi}$ by removing $\theta_{jq}$, $\nu_{jq}$, and $\tau_{jq}$
% \If{Non-unique $\theta_{jq}$ found}
%     \State $d_{\alpha+1}=d_{\alpha}/2$
%     \State $\alpha=\alpha+1$
%     \State Go back to step~\ref{step2}
% \EndIf
% \EndWhile
% \end{algorithmic}
% \end{algorithm}

% \medskip
% \noindent\textit{Communication performance ($N_t=4$, $N_c=8$)} - We evaluate the BER of the proposed system with  different number of private bins. 
% % The~\cref{fig:private_communication} also includes results of the all-shared bin design for comparison. 
% When $N_c=8$, we observe that the all-shared bin design retains the result presented in~\cref{fig:comm}.
% Additionally, we observe the BER of OTFS is not affected under different number of private bins.
% This is because the proposed zero enforcement avoid DD signal distortion.
% So the DD signal can be recovered by~\cref{21}.

% \textcolor{red}{
% The sensing performance advantage of the virtual array occurs at the cost of communication rate.
% Here we demonstrate the BER performance versus the number of private bins $P$ for different SNR levels. *****
% *****show results for $P=2,4,6$. The horizontal axis should be SNR -10 to 30 db.
% Create a new  graph for Probability of target detection for the same P and SNR.
% %
% For the probability of target detection do monte carlo simulations; in each run you have targets randomly placed on the target grid, separated by angle distance greater that $x$ degrees. Vary $x$ so that we can see the limit.
% }

\subsection{Communication of the proposed system}
\begin{figure}
    \centering
    \includegraphics[width=0.7\linewidth]{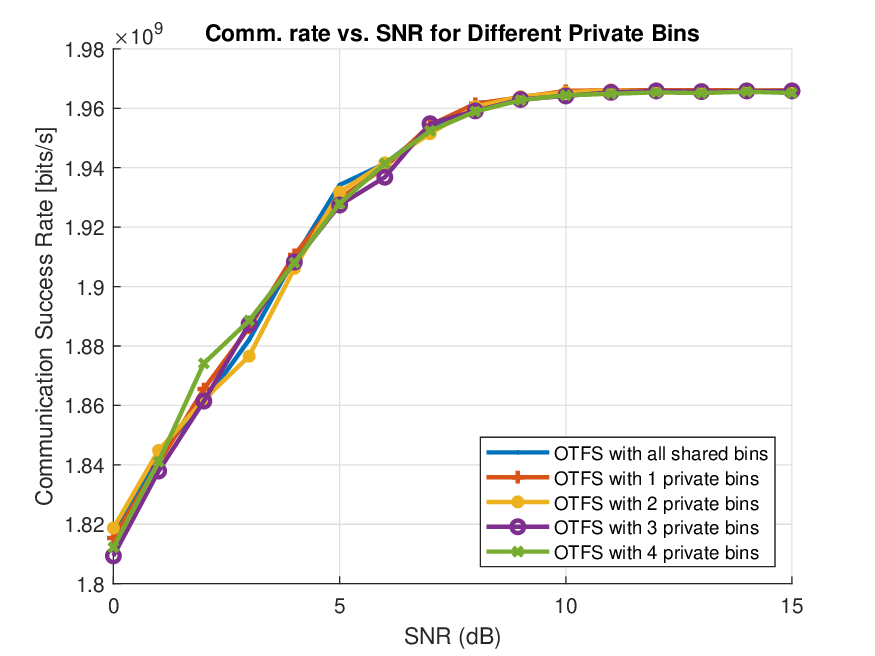}
    \caption{Communication Rate Loss Using Private Bins.}
    \label{fig:comm_rate}
\end{figure}

\textcolor{black}{
We evaluated the communication  rate of the proposed system with different numbers of private bins under different SNRs.
For the case of $N_t=4$, $N_c=8$, as shown in fig.~\ref{fig:comm_rate}, the private bin design has similar communication performance compared to the all shared bin design. 
}%
The loss of communication rate due to the use of private bins is rather small.
With $N_t=4$ and using QPSK symbols, each private bin leads to $3\times \log_2(4)\times 120\mathrm{kHz}=7.2\times 10^5$ [bits/s] communication loss.
When $4$ private bins are used, the percentage communication rate loss is $\frac{N_t\times (4-1)}{N_t\times NM}=0.037\%$.

\section{Conclusion}\label{sec:conclusion}
We have proposed a novel DFRC MIMO OTFS system that is robust to Doppler shifts, can efficiently use the bandwidth for communication and sensing, and is equipped with a low-complexity high-resolution radar parameter estimator.
DD domain bins are used in a shared fashion, while some TF bins are private to a small number of transmit antennas. 
The shared bins enable high communication rates. 
Introducing private bins requires some reduction of the DD domain symbols transmitted by each antenna. However, it 
enables the formation of a virtual array that improves the sensing performance. 
A small number of private bins, or equivalently, a very small amount of rate loss, suffices to achieve significant sensing gains.

% \textcolor{red}{use consistent format in references - check names }

% \bigskip
% \textcolor{magenta}{I see inconsistencies in the bibliography. I cannot bring this up in every paper. Please fix}
% \textcolor{blue}{fixed}

\bibliographystyle{IEEEbib}
\bibliography{bib}

\end{document}